\documentclass[journal]{IEEEtran}
\ifCLASSINFOpdf
  \usepackage[pdftex]{graphicx}
\else
\fi
\usepackage{amsmath}
\usepackage{algorithm}
\usepackage{algpseudocode} 
\usepackage{xcolor}

\usepackage{array}
\usepackage{tabularx}
\begin{document}
%
\title{Directional Antenna Based Scheduling Protocol for IoT Networks}
%
%
%

\author{
\IEEEauthorblockN{Anil Carie\IEEEauthorrefmark{1}, Abdur Rashid Sangi\IEEEauthorrefmark{2}\IEEEauthorrefmark{3}, Satish Anamalamudi\IEEEauthorrefmark{1}, Murali Krishna Enduri\IEEEauthorrefmark{1}, Baha Ihnaini \IEEEauthorrefmark{2}\IEEEauthorrefmark{3}, Hemn Barzan Abdalla\IEEEauthorrefmark{2}\IEEEauthorrefmark{3}
}
    \\\
    \IEEEauthorblockA{\IEEEauthorrefmark{1}SRM University-AP, India.
    }
    \IEEEauthorblockA{\IEEEauthorrefmark{2}Wenzhou-Kean University, P. R. China.
    }
    \IEEEauthorblockA{\IEEEauthorrefmark{3}Kean University, NJ, USA.
    }
\thanks{Anil Carie is with the Department
of Computer Science and Engineering, SRM University-AP, Amaravati, AP, India. email: carieanil@gmail.com}
\thanks{Abdur Rashid Sangi  is with the Department of Computer Science, College of Science and Technology, Wenzhou-Kean University
, 88 Daxue Road, Ouhai, Wenzhou, Zhejiang 325060, China. email: sangi\_bahrian@yahoo.com}
\thanks{Satish Anamalamudi is with the Department
of Computer Science and Engineering, SRM University-AP, Amaravati, AP, India.
 e-mail:satishnaidu80@gmail.com }
 \thanks{Murali Krishna Enduri is with the Department
of Computer Science and Engineering, SRM University-AP, Amaravati, AP, India.
 e-mail:muralikrishna.e@srmap.edu.in }
 \thanks{Baha Ihnaini is with the Department of Computer Science, College of Science and Technology, Wenzhou-Kean University
, 88 Daxue Road, Ouhai, Wenzhou, Zhejiang 325060, China. email: bihnaini@wku.edu.cn}
\thanks{Hemn Barzan Abdalla is with the Department of Computer Science, College of Science and Technology, Wenzhou-Kean University
, 88 Daxue Road, Ouhai, Wenzhou, Zhejiang 325060, China. email: habdalla@kean.edu}
\thanks{Manuscript received month date, year; revised Month date, year.}}

%
%

\markboth{Journal of \LaTeX\ Class Files,~Vol.~14, No.~8, Dec~2021}%
{Shell \MakeLowercase{\textit{et al.}}: Bare Demo of IEEEtran.cls for IEEE Journals}
%



\maketitle

\begin{abstract}
Scheduling and Channel Access at the MAC layer of the IoT network plays a pivotal role in enhancing the performance of IoT networks. State-of-the-art Omni-directional antenna based application data transmission has relatively less achievable throughput in comparison with directional antenna based scheduling protocols.
To enhance the performance of the IoT networks, this paper propose a distributed one-hop scheduling algorithm called Directional Scheduling protocol for constrained deterministic 6TiSCH-IoT network. With this, increased number of IoT nodes can have concurrent application data transmission with  efficient spatial reuse. This in-turn results in higher number of cell allocation to the one-hop IoT nodes during data transmission. The proposed algorithm  makes use of through directional transmissions avoids head of line blocking.

\end{abstract}


%
\IEEEpeerreviewmaketitle

\section{Introduction}
%
%
%
%
\IEEEPARstart{T}{he} Internet of Things (IoT) has the potential to meet the demands of rapidly evolving wireless technologies, each with distinct infrastructural needs. IoT facilitates the connection of heterogeneous networks (such as Wi-Fi, Bluetooth, etc.) to the internet, creating a new paradigm for information analysis and decision-making. 

The IEEE 802.15.4e \cite{ieee2011802} is being proposed as a protocol for IoT, defining the MAC and physical layers of Low-power and Lossy Networks (LLNs), which have facilitated increased IoT deployment. One of the five MAC operation modes in the IEEE 802.15.4e protocol is Time Slotted Channel Hopping (TiSCH). In TiSCH, all IoT nodes use a common schedule to determine the time slot and channel for communication with their one-hop neighbors. However, the standard does not specify how the schedule should be constructed and updated. As IoT becomes integrated into everyday life, IoT-enabled nodes generate high volumes of heterogeneous traffic. The routing and scheduling implemented in this standard directly affect node energy consumption and propagation delay. 


Due to the resource constraints of sensor nodes, the Internet Engineering Task Force (IETF) is developing various efficient low-energy communication standards. The IETF 6TiSCH (Time Slotted Channel Hopping) working group \cite{wang20156tisch} outlines a mechanism that combines the high reliability and low energy consumption of IEEE 802.15.4e TiSCH with the IP protocol. This integration ensures interoperability and smooth application data transmission. The 6TiSCH Operation Sublayer (6top) defines a scheduling function (SF0) that adds or deletes cells between neighboring IoT nodes by monitoring and collecting user data. However, the current implementation of the Scheduling Function in 6TiSCH operates in an omni-directional mode.

Using directional antennas can significantly reduce power consumption and minimize one-hop interference with neighboring nodes. Additionally, multiple IoT nodes within the same coverage area can communicate more effectively by enhancing spatial reuse. However, to fully leverage directional transmission, an efficient medium access control protocol must be designed to address issues such as deafness and hidden terminal problems.  

The 6TiSCH nodes are synchronized based on common frame structure along with transmission schedule. However, the key problem with the IEEE 802.15.4e is that it doesn't specify the time slot scheduling protocol. In this work, a directional antenna based distributed scheduling is being proposed where every IoT node prepares its schedule in distributed mode. Since sensor nodes exchange the data frequently, efficient MAC design should be proposed to allocate the time slot for the node intending for communication. Our protocol uses directional antenna for sending and receiving the application data in between one-hop neighbor IoT nodes. This needs to have a strict node synchronization during the cell schedule with its neighbor. To address the issue, this paper start with the discussion of use of directional antenna in 6TiSCH network to reduce the interference and increase the spatial reuse. 
Later, distributed scheduling strategy is explored to determine the available slots when a node needs to either send or receive messages from its one-hop IoT neighbor nodes. The main contribution of this proposed work are as follows:

\begin{itemize}
    \item A Hybrid TiSCH MAC for IoT nodes, to calculate global schedule for their transmission in distributed manner.
    \item Use of Omni-Directional antenna for control message and Directional antenna for data transfer.
\end{itemize}

The rest of paper is organized as follows. Section II presents about the brief review of literature on IoT enabled MAC protocols and 6TiSCH scheduling algorithms. Section III describes about the system model, antenna model, assumptions and the proposed work. Section IV explains about the simulation results and analysis of the proposed model. Finally, Section V is about conclusion and  insights of the future work.

\section{Related Work}

In the following two sub-sections, the focus is on existing energy conservation techniques and the scheduling algorithm for the IoT networks. 

\subsubsection{IoT MAC Protocols}
In \cite{ye2017token}, TA-MAC is proposed for IoT enabled mobile ad hoc network(MANET) where nodes are grouped into one hop neighbors and TDMA super frame is used to allocate time to different groups avoiding hidden terminal problem. A distributed token passing mechanism is used to allot time slots to nodes in each group. MAC parameters are optimized to minimize end to end delay.
In \cite{ma2018design}, issue of centralized MAC is addressed where large number of back-scatter devices may join/leave the network causing frequent topology changes making it hard to maintain devices in dynamic network. Thus, distributed medium access control(MAC) is devised for co-existence of Wi-Fi and backscatter communication.
In \cite{ye2016distributed}, distributed and adaptive MAC for 1-hop IoT enabled MANET is proposed where voice packets are transmitted using contention free channel access using distributed time division multiple access while data is transmitted over contention based channel access using carrier sense multiple access with collision avoidance. Voice capacity is increased using multiplexing while data throughput is maximised by adjusting the optimal contention window.
Authors proposed MAC paradigm called EMIT to address the high communication cost of the large number of IoT devices \cite{bakshi2019emit} by adopting interference averaging strategy where nodes share the available resources concurrently. wireless local area networks generate large volume of traffic from sparsely positioned stations while traffic generated by densely deployed IoT devices is sparse in time.
As the number IoT devices is huge this will result in congestion which increases number of re-transmissions and power consumption. To address this an efficient hybrid MAC is proposed in \cite{al2019energy} which uses the available energy efficiently and adapts to sleep and awake periods corresponding to network load. 
In \cite{iqbal2019gwins} authors addressed the energy constraints in IoT devices through energy harvesting and data relaying in distributed manner using hybrid access points and to reduce the collisions a group based MAC protocol is proposed. 
Authors in \cite{cao2018performance} designed hybrid MAC where contention phase is succeeded by multi-slot reservation transmission(MRT) for efficient transmission. Multiple slots are reserved by node pairs before the contention phase and data transmission occurs in MRT. However, the number of node pairs is assumed in this work which is a vital factor and superframe length is fixed. 
Researchers \cite{liu2014design} have focused on hybrid MAC protocols( CSMA/CA and TDMA) for IoT devices due to high collision rate using carrier sense multiple access/collision avoidance(CSMA/CA) and less slot utilization TDMA. Most of the existing work related to IoT enabled network is with the primary assumption that every node is equipped with omni-directional antenna. Since the signal is radiated in all directions,nodes suffer with interference from specific directions and short transmissions range.

\subsubsection{Scheduling Algorithms for 6TiSCH}
The existing literature on 6TiSCH scheduler \cite{phung2018scheduler,10284585} is extensive and focus particularly on centralized and distributed approaches. In centralized approach nodes need to have prior information of topology and traffic, then scheduler will assign slots to all the nodes based on the requirements. However, in network where the typologies change dynamically and traffic is sporadic centralized mechanisms are slow to adapt as the scheduling algorithm needs to run for the new information. On the other hand, distributed approaches are easy to adjust to the network changes ( topology, traffic , channel variations), however to reaching optimal solution is subject to the efficiency of the algorithm. In this solution, every node prepares its own schedule based on the local information available.
In \cite{palattella2015fly}, on-the-fly (OTF) distributed algorithm is used for allocation of bandwidth where the number of cells are added/deleted based on the availability and the requirement, for successful transmission of the data. This approach is useful in event detection models where the information need be sent to central node (sink) quickly. However, this method suffers with collision as the scheduling is done based on local information unaware cells allocated to other pair of nodes.
In \cite{palattella2015fly}, authors classify the node traffic based on the data rate required and a schedule is made by super positioning 3 slot frames. Every node calculates hash values based on its MAC to determine which time slot to use. This approach has less communication overhead but its suffers from high latency.
Decentralized Broadcast Scheduling Algorithm (DeBraS) \cite{municio2016decentralized,9745528}, share the local schedules and avoids collisions, the cost of reducing collisions in dense networks is at the expense of higher energy consumption 


\newcommand{\tabincell}[2]{\begin{tabular}{@{}#1@{}}#2\end{tabular}}
\begin{table}[!t]
  \scriptsize
  \caption{Notations}
  \label{tab:notations}
  \begin{tabular}{ll}
    \\[-2mm]
    \hline
    \hline\\[-2mm]
    {\bf \small Symbol}&\qquad {\bf\small Meaning}\\
    \hline
    \vspace{1mm}\\[-3mm]
    $N$      &   \tabincell{l}{number of nodes in the network }\\
    \vspace{1mm}
    $n_i$          &  \tabincell{l}{$i^{th}$ node in the network}\\
     \vspace{1mm}
    $V$          &  \tabincell{l}{$V={n_0,n_1,....n_N}$ are the sensor nodes with $n_o$   as the \\root node }\\
     \vspace{1mm}
     $R_i$          &  \tabincell{l}{radius of communication}\\
     \vspace{1mm}
    $D_{x,y}$          &  \tabincell{l}{Direction of the node}\\
    \hline
    \hline
  \end{tabular}
\end{table}


\section{System Model}
\subsection{Antenna Model}

In this work, switched beam antenna is used, as it supports fixed and highly directive beam for data transmission. The beam width $\theta$ is set to 90$^{\circ}$ to point in four directions for simplicity, however it can be changed to M directions. An IoT node switches from one beam to another visiting all the beams is defined as sweeping. Neighbor node position is evaluated based on the direction of arrival(DOA). All the nodes in the network are equipped with directional antennas.Thus, the proposed scheduling protocol can achieve reduced node energy consumption with maximizing the concurrent transmissions with minimal interference. It is noteworthy that each and every IoT node will maintain cell-allocation matrix(to keep track of cell information) and directional-antenna beam matrix.

IoT Nodes(Sender) start scheduling with its one hop neighbor nodes based on local information (cell availability ) and broadcast its schedule information to local neighbours. neighbor Nodes that receive schedule information from sender will update in their cell-allocation matrix and directional beam matrix.With this, nodes who are in the communication range of the sender will know about the antenna beam that is going to be used for directional transmission.

In current WSN hardware is each node is equipped with half-duplex transceiver. This leads to two categories of conflicts: (i) primary and (ii) secondary. When transmit and receive simultaneously primary conflict and when node receives multiple transmissions causes secondary transmission \cite{nirapai2014centralized} .

\begin{figure}[!t]
\centering
\includegraphics[width=3.4in]{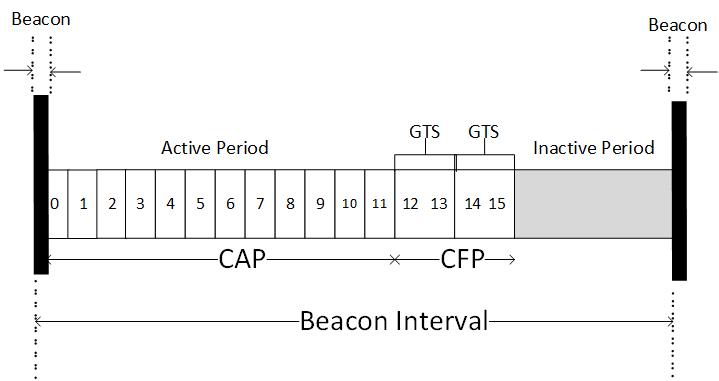}
\caption{Superframe structure of IEEE 802.15.4}
\label{superframe}
\end{figure}

For secondary conflict, let us assume that sink location is fixed and sensor nodes is deployed within the IoT network. Once the sensor nodes is battery powered, how exactly the synchronization happens among the one-hop neighbor nodes is a challenging task.
To achieve this, nodes that are switched ON will listen in all directions to receive the scheduling update from the one-hop neighbor nodes. Later, data transmission will get broadcast with the directional antenna towards direction of the sink. 

\subsection{Communication and Computation Delays }
In the proposed model, the minimum rate of upload between two nodes i and j is given by ${ru}_{i,j}$, maximum delay at any given node is defined as

\begin{equation}
  D^u = max\left(\frac{|B_i|}{ru_{i,j}} + \zeta_i^u\right) \forall i \in V
  \label{nactive}
\end{equation}
where $|B_i|$ is the total number of bits to be transmitted in up-link and down-link at a given node, $\zeta_i^u$ is the channel access time. Similarly, the minimum rate of download between two nodes j and j is given by  ${rd}_{j,i}$, maximum delay at any given node is defined as 
\begin{equation}
  D^d = max\left(\frac{|B_i|}{rd_{i,j}} + \zeta_i^d\right) \forall i \in V
  \label{nactive}
\end{equation}

The computation time per iteration at node depends on the size of the learning model, CPU cycles to execute one unit of data is denoted by $\chi_i$, and the size of dataset $S_i$. The number of CPU cycles needed to execute is $\left(\chi_i . S_i\right)$, given all the data.

\subsection{Directional Energy Consumption}
The energy consumption at node $i$ is function of channel state, directional transmission rate and the bandwidth. The wireless channel between an nodes i and j is given by signal to noise ratio(SNR) $\gamma_{ij}$ defined as 

\begin{equation}
    \gamma_{ij} = \frac{P_{ij}^r . |h{ij}|^2}{N_0.W_{ij}},
\end{equation}
where $P_{ij}^r$, is the received power at i, $|h{ij}|$ is the channel fading, $N_0$ is spectral density and $W_{ij}$ is the bandwidth. Also, the path loss for distance of $d_{ij}$ is $\alpha$ $\left(2 \leq\alpha \leq 6\right)$. Thus received power is attenuated with respect to transmission as follows $P_{ij}^r=P_{ij}^t . \omega . d_{ij}^{-\alpha} $. Transmission rate at node i to j given SNR is defined as 

\begin{equation}
    ru_{i,j}= W_{ij}\log_2(1 +\upsilon \gamma_{ij})
\end{equation}
where $\upsilon = -1.5/ \log(5. BER)$ BER is the bit error rate, the transmitted power can be written as

\begin{equation}
    P_{ij}^t = \frac{N_0 . W_{ij}}{g_{ij}}\left( 2^\frac{r_{ij}}{W_{ij}} -1\right)
\end{equation}
Where $g_{ij}$ is the channel gain that is defined as \begin{equation}
    g_{ij} = \upsilon .\omega .d_{ij}^{-\alpha}. |h_{ij}|^2
\end{equation}

Total energy consumed at node i to send a data of length $B_i$ to node j is

\begin{equation}
    E_{ij}= \frac{P_{ij}^t . |B_i|}{r{ij}}
\end{equation}

\subsection{Design Objectives}
\textbf{Minimal Schedule Length}
The primary design objective is to make use of concurrent transmissions, increasing the spatial reuse by minimizing the interference to reduce the schedule length.

\textbf{Minimizing Energy Consumption:}
Two popular techniques for maximizing the network life time in resource constrained WSN are:(i) transmission power control and (ii) radio activity. Nodes can transmit with optimal power instead of maximum power and instead of keeping the radio active all the time, intelligent scheduling  between sleep and active stat is archived using duty-cycles.

\subsection{Network Model}
In TSCH, enhanced beacons(EBs) are broadcast by coordinator for network formation. Then, interested nodes scan for the EBs messages and replies with EBs message show its presence.In this model we consider a 6TiSCH network with tree topology built using Routing Protocol for Low-Power and Lossy Network. We consider a time-slotted IEEE 802.15.4-2015 networks, where the nodes operate in a distributed manner. The network can be defined as $G = (V,E)$ where $V={n_0,n_1,....n_N}$ is the set of nodes in the network. Here, $n_0$ is the sink, $E$ denotes the link between the nodes. $R_i$ denotes the communication radius, $D_{x,y}$ denotes the direction of the node. Sensor nodes monitor various events and forwards it to sink node.
In the proposed work, antenna works in directional mode for \textcolor{red}{data transmission} and omni-directional mode for data collection. \textbf{Node directional matrix} avoids interference from neighbor nodes accessing the same channel. Every nodes knows its own location, by the exchange of local information  node calculates the antenna direction and index number of neighbor nodes using global position system (GPS).

Directional interference in a channel is avoided through modified RTS and CTS message which include index of antenna and angle of arrival(AoA) information. Due to this directional data transfer with efficient transmit power is archived.

\textbf{Node Channel Matrix}

The primary objective of node-channel matrix is to avoid collisions and reduce power consumption.
\begin{equation}
M=\begin{bmatrix}
V_{[1,1]}^{[K]} & \cdots & V_{[1,C^{[1]}]}^{[K]}\\
\vdots & \ddots & \vdots\\
V_{[K,1]}^{[K]} & \cdots & V_{[K,C^{[K]}]}^{[K]}
\end{bmatrix}
\end{equation}

\textbf{Node Directional Matrix}
We design node-directional matrix (2) to avoid interference with neighbor directional data transmission on same channel.

\begin{equation}
M=\begin{bmatrix}
D_{1,2} & \cdots & D_{1,K}\\
\vdots & \ddots & \vdots\\
D_{K,1} & \cdots & D_{K,K-1}
\end{bmatrix}
\end{equation}

\begin{figure}[!t]
\centering
\includegraphics[width=2.5in]{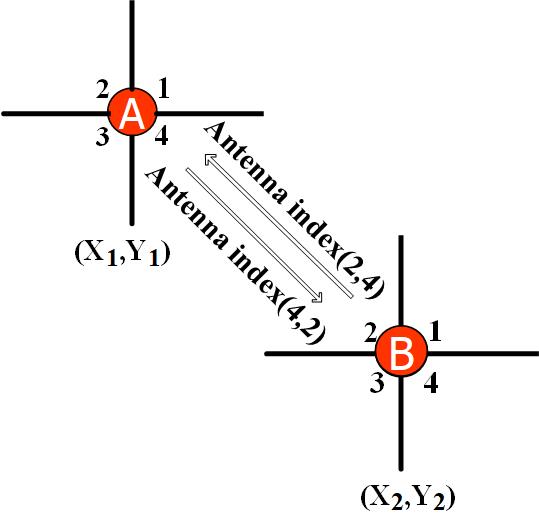}
\caption{Directional antenna discovery for neighbour node}
\label{directional}
\end{figure}

\subsection{Directional Transmission}

\begin{figure}[!t]
\centering
\includegraphics[scale=.5]{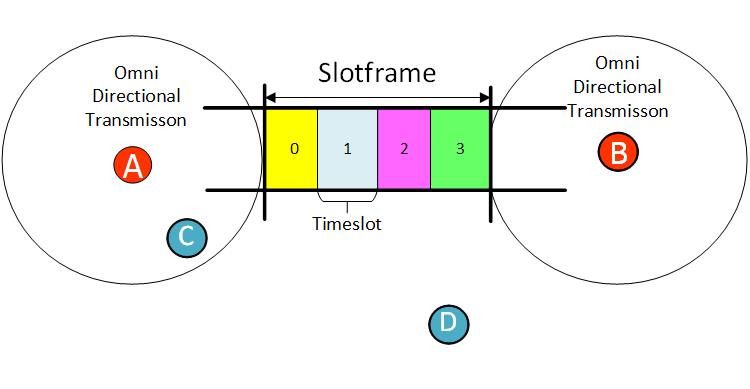}
\caption{Omni directional TSCH transmission from node A to node B, node C and node D are in waiting state}
\label{OTSCH}
\end{figure}

In the above scenario figure \ref{OTSCH}, we assume 4 nodes {A,B,C,D} using one channel trying to have communication with each other, node A want to communicate with node B , also node C wants to communicate with node D. Node pair (A,B) is in need of 4-time slots and is scheduled to transmit in channel 1's four time slots starts its omni directional transmission while node pair (C,D) have to wait for next cycle for transmission. Omni directional transmission prevents surrounding nodes which are in its interference range from transmitting. With the use directional transmission, node C which is intending to transmit in other direction is allowed start its transmission. Thus, in directional transmission nodes both the node pair can access the channel simultaneously increasing the spatial reuse, reduce the waiting time.

\begin{figure}[!t]
\centering
\includegraphics[scale=.58]{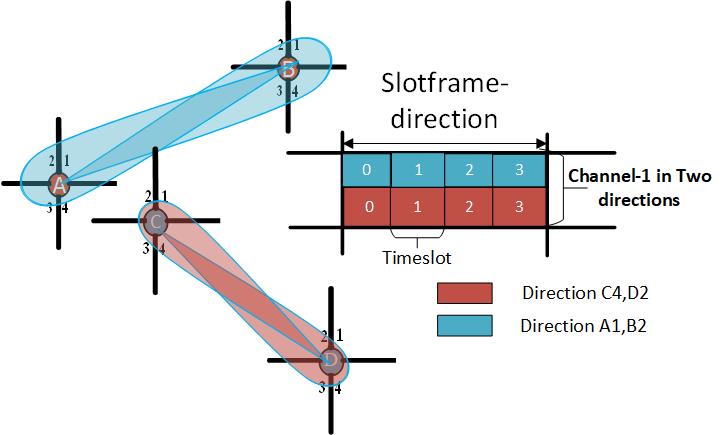}
\caption{Directional TSCH transmission form node A to node B and node C to node D}
\label{DTSCH}
\end{figure}

\subsection{Node Scheduling}
Data delivery rate is constrained by the rate at which sink can accept the data. High priority is given to 1-hop neighbors as they are busy all the time. The scheduling of 1-hop neighbours is adjusted based on the packet arrival rates. A top-subtree TS(r) is defined as one whose root $r$ is a child of the sink, and it is said to be eligible if $r$ has at least one packet to send. Any given node has to identify the slots through which it can send information and the slot which it can receive information, for the selected time slot node needs to decide channel offset. Nodes around the sink prioritized based on the min buffer size and high available bandwidth. Nodes for which sink is in directional transmission range have two modes of operation.

\section{The Proposed MAC Model}
Hybrid MAC with synchronous TDMA-CSMA/CA is designed to schedule the sensor nodes for directional data transmission.

\subsection{Synchronous Scheduling}
\begin{figure}[!t]
\centering
\includegraphics[scale=.5]{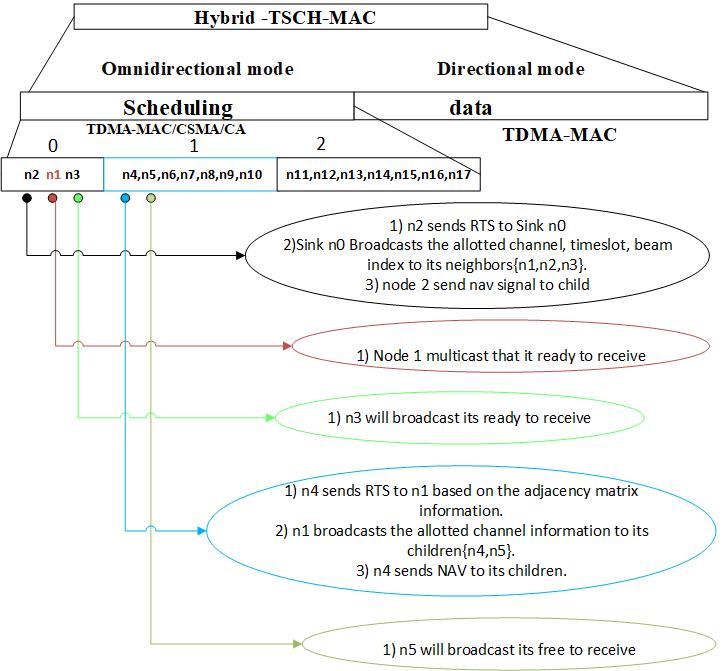}
\caption{Level Based Scheduling}
\label{scheduling}
\end{figure}

\begin{figure}[!t]
\centering
\includegraphics[scale=.5]{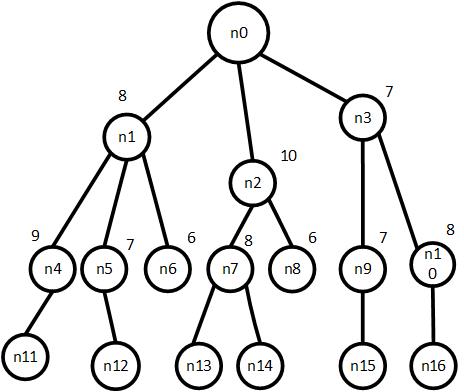}
\caption{Top Subtrees with varying loads}
\label{tree}
\end{figure}

Our proposed model is based on tree topology consisting of one parent node and multiple child nodes as shown in figure \ref{tree}. The protocol includes two periods, synchronous scheduling  and asynchronous data transfer. Our main idea is to schedule nodes based on the timer value which is calculated based on the number of packets remaining (buffer size). Since, many-to-one communication employed in sensor nodes suffers from a congestion problem called the funnelling effect. So, one of the Primary objectives is to keep the sink node busy. The synchronous scheduling period is divided into there TDMA slots ${0,1,2}$ of equal length, in slot-0 nodes at level-1 are scheduled to start the timer. In the \ref{scheduling}, node $n2$ timer expires first, hence it sends RTS to root node $n0$. Next, $n0$ broadcasts CTS to its neighbour nodes ${n1,n2,n3}$, upon receiving CTS from the root node $n0$, $n2$ sends a NAV signal to its child nodes indicating that it is sending mode. Next, node $n1$ timer expires it has listened to the CTS for $n2$ node from $n0$, so it changes its mode to receiving mode and broadcast to its neighbors it availability. Similarly node $n3$, changes to receiving mode and informs its availability to its children. As, the slot-1 begins all the nodes level-2 will start their timer, $n4$ timer expires first and its sends are RTS to $n1$ node which is in receiving mode. Node $n2$ sends are CTS to all its children ${n4,n5}$, $n5$ understand $n1$ is busy and changes its mode to receiving mode. Similar, procedure followed for all the remaining node the till all the possible links are scheduled.

\begin{algorithm}[htb] 
\caption{Distributed Scheduling} 
\label{alg:Framwork} 
\begin{algorithmic}[1] 
\Require 
Every node knows its level in the tree, and are fed with timer values when they need to start their timers.
\label{Initilization} 
\State Nodes start their timer when the scheduling slot for their level begins.
\State  Initially, 1-hop neighbors from the sink start a timer(based on the number of packets remaining); 
\label{schedule} 
\State Node whose timer expires is assigned a slot-channel pair(towards sink) in the schedule and is updated in the schedule. 
\State Parent node share the scheduled link information with child node (i.e. sink will broadcast the scheduled link information to node 1 and node 3. 

\State when n1 timer expires node 1  will broadcast to its children and  similarly, node 3.
\label{update} 
\State All the nodes update the information in their adjacency tables.
\State At the beginning of second time slot, all two hop neighbors start their timer.
\State node 4 will check its adjacency table, and send request to n1 which is in receiving mode. n1 will multicast confirmation request. n5,n6 will update their table. 
\State Nodes not violating the adjacency constraint can be scheduled in parallel.

\label{code:fram:classify} 
\State Child nodes reserve link based on the information received form the parent node. 
\end{algorithmic} 
\end{algorithm}



\subsection{Single Channel Omni-Directional Case}
\begin{figure}[!t]
\centering
\includegraphics[scale=.5]{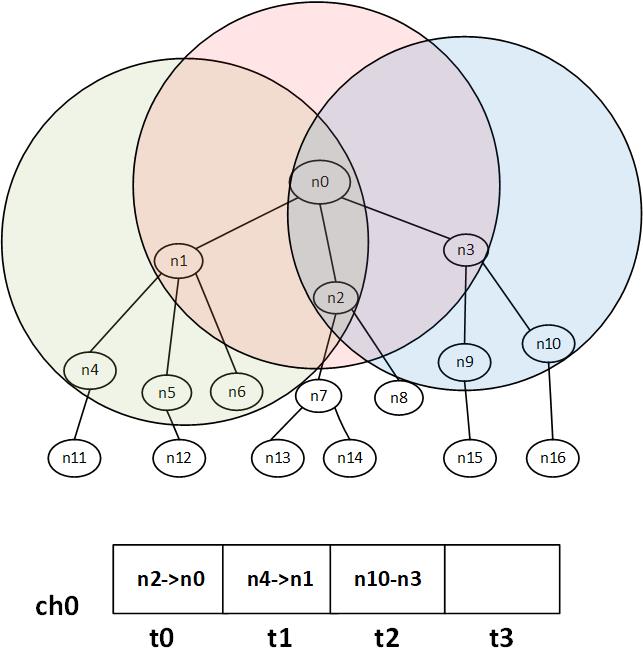}
\caption{Single Channel Omni-Directional Scheduling}
\label{SCO}
\end{figure}

In this scenario as shown in the figure \ref{SCO}, nodes $ n0,n1,n2,n3$ are in the interference range of one another and they are operating with omni-directional antenna with one available channel. Thus, when node $n2$ is transmitting in channel-1 and time slot-0, neighbor nodes $n1,n3$ need to be silent in time slot-0. Transmissions between n4->n1 and n10->n3 need to be scheduled in different time slots, in order to prevent interference to node $n2$.

\subsection{Single Channel Directional Case}
\begin{figure}[!t]
\centering
\includegraphics[scale=.5]{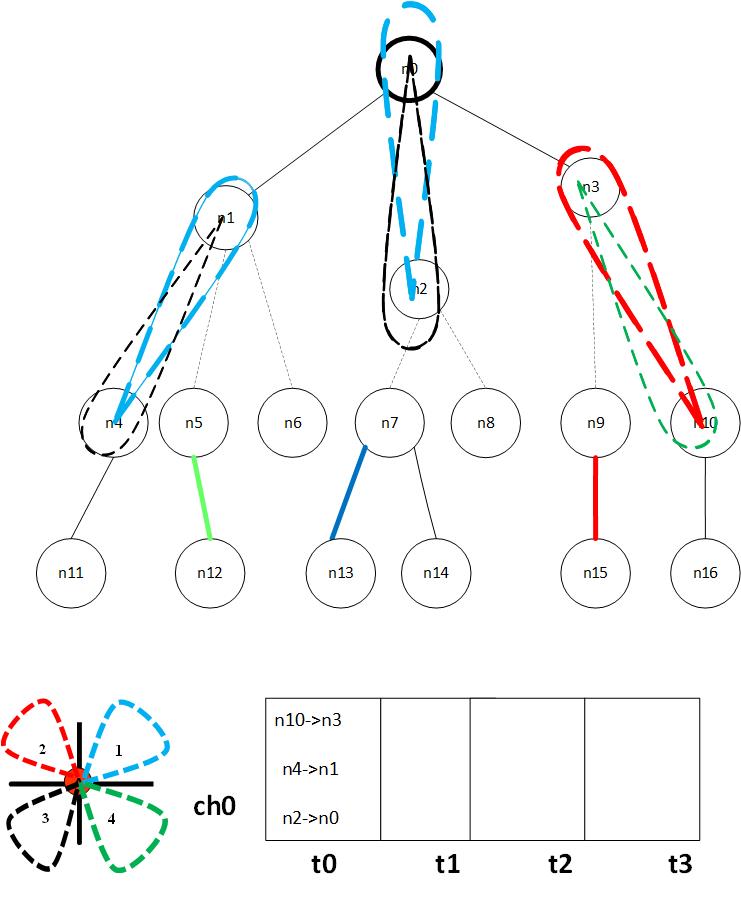}
\caption{Single Channel Directional Scheduling}
\label{SCD}
\end{figure}

With the use of directional transmission as shown in figure \ref{SCD}, nodes which are in interference range of each other i.e, $n0,n1,n2, n3$ can be scheduled to operate in single channel, in different directions. Transmission between $n2->n0$ uses (1,3) beam to communicate, $n4->n1$ uses (1,3) and $n10->n3$uses (2,4) beams respectively. All the three transmissions are scheduled in channel-0, time slot-0 simultaneously. Thus spatial reuse achieved improving, throughput, reducing delay and energy consumption. Hence, the proposed model helps to realize the minimal schedule length and minimal energy consumption.

\section{SIMULATION SETUP}
For simulations, we consider a wireless IoT network that consists of N = 16 nodes one of them is sink node, the example network shown in fig. \ref{tree}. The $n-1$ nodes are are uniformly distributed and the distance between node i and the sink is denoted as $d_i$. Every nodes wireless channels follow i.i.d. Rayleigh fading with the total allowed bandwidth $W_{ij}$ = 20 MHz, is modeled as $h_{iJ}$. We have used quasi-switched beam antenna which can steered with a span of 360 degrees. Parameters for simulation are shown below Table \ref{tab:my-table}

\begin{table}[]
\centering
\caption{Parameters used in Simulation}
\label{tab:my-table}
\begin{tabular}{|l|l|}
\hline
\textbf{Parameters} & \textbf{Value} \\ \hline
Area                & 1000 x1000 m   \\ \hline
Number of nodes     & 16             \\ \hline
Transmission Power  & 15 dBm         \\ \hline
Receiving Threshold & -81.0 dBm      \\ \hline
Sensing Threshold   & -91.0 dBm      \\ \hline
Data Rate           & 2Mbps          \\ \hline
Packet Size         & 127 bytes      \\ \hline
Simulation Time     & 5 minutes      \\ \hline
\end{tabular}
\end{table}

\section{RESULT ANALYSIS and DISCUSSION}
\begin{figure}[!t]
\centering
\includegraphics[width=3in]{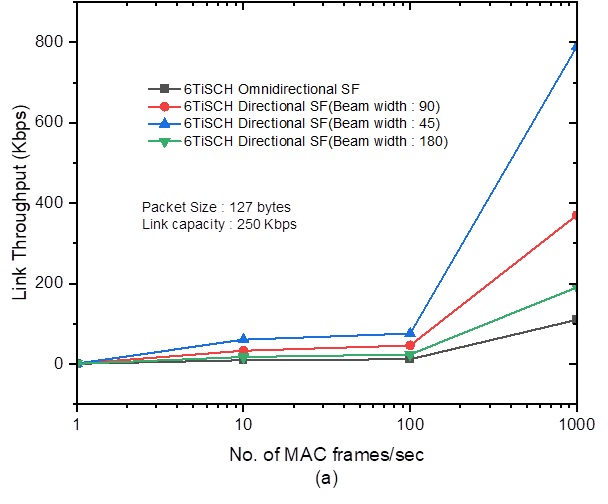}
\caption{Directional antenna discovery for neighbour node}
\label{link}
\end{figure}

\begin{figure}[!t]
\centering
\includegraphics[width=3in]{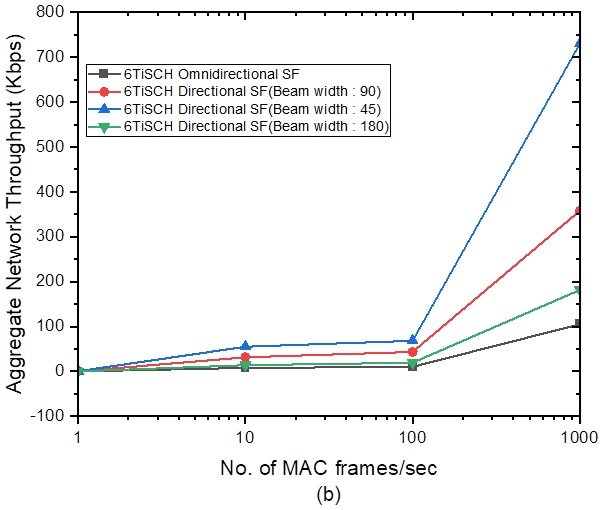}
\caption{Directional antenna discovery for neighbour node}
\label{directional2}
\end{figure}

\begin{figure}[!t]
\centering
\includegraphics[width=3in]{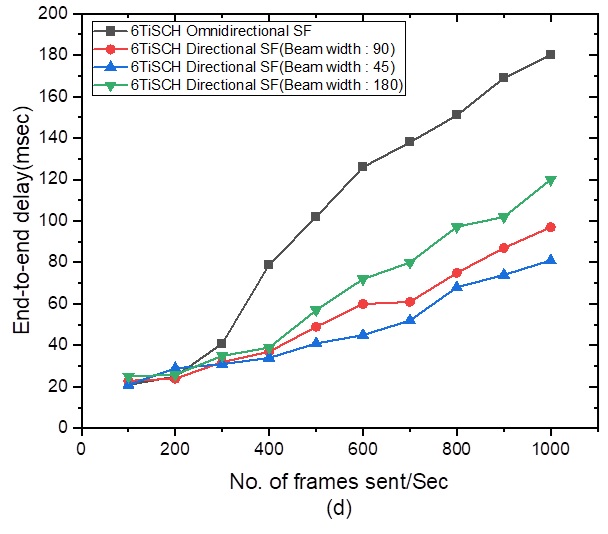}
\caption{Directional antenna discovery for neighbour node}
\label{directional4}
\end{figure}

We have simulated IEEE 802.15e time slotted channel hopping with directional antennas\cite{10525788}. The average link, network throughput and end-to-end delay are evaluated with increasing beam width in Fig \ref{link}, Fig. \ref{directional2} and Fig. \ref{directional4}.

In Fig. \ref{link}, as the date rate is increasing, link throughput of our proposed 6TiSCH with directional antenna is much better than that of 6TiSCH with Omni-directional SF. Also, average End-to-End delay is approximately halved that that of 6TiSCH omni directional.
Similarly, as you can see in the Fig. \ref{directional2}, the average network throughput of the proposed protocol is much better as the nodes which don't chance to send data to sink node, are not idle, but they collect information form their child nodes.

\begin{figure}[!t]
\centering
\includegraphics[width=3in]{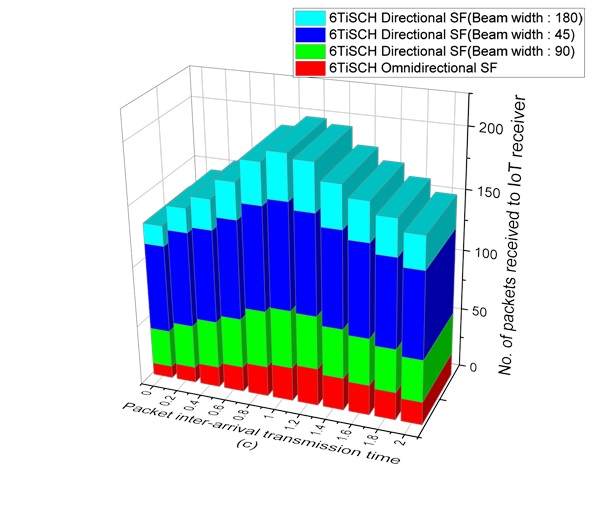}
\caption{Directional antenna discovery for neighbour node}
\label{directional3}
\end{figure}

Also, in Fig. \ref{directional3} and Fig. \ref{directional4} the End-to-End delay is significantly decreases in the proposed directional protocol. Since, every node if fails to find time-slot to send data to parent node, announces its availability to receive data packets from child.

While the proposed Directional Scheduling protocol shows promising results in enhancing the performance of constrained deterministic 6TiSCH-IoT networks, there are still areas that can be further improved. One drawback of the proposed algorithm is that it relies on directional transmissions, which may not be feasible in certain scenarios where nodes have limited mobility or fixed locations. In addition, the protocol currently focuses on one-hop scheduling and does not address multi-hop scenarios, which may be relevant for larger IoT networks.

Future research can explore the use of machine learning and artificial intelligence (AI) techniques to optimize the scheduling and channel access at the MAC layer of IoT networks. AI can be used to learn from the network environment and make intelligent decisions on scheduling and resource allocation based on various factors such as network congestion, channel quality, and mobility patterns of the nodes. This can potentially improve the performance of IoT networks in various scenarios, including those with limited resources and mobility constraints.

Furthermore, future work can investigate the integration of the proposed Directional Scheduling protocol with other existing scheduling protocols, such as Time Division Multiple Access (TDMA) and Carrier Sense Multiple Access with Collision Avoidance (CSMA/CA), to further improve the performance of IoT networks.

\section{Conclusion}
In this paper, we have proposed distributed scheduling for IEEE 802.15.4e network in Time slotted channel hopping (TSCH) MAC mode. With the use of directional antenna has helped to reduce the power consumption, collisions, delay while increasing the spatial reuse and parallel transmissions. 
To address these limitations and further enhance the performance of IoT networks, future research could explore the use of AI and blockchain technologies. AI can be used to dynamically adjust the scheduling algorithm to account for changing network conditions, while blockchain can ensure secure and efficient communication between IoT devices. Additionally, future work could investigate the use of hybrid directional and omnidirectional antenna-based scheduling protocols to achieve higher throughput without the cost of complex hardware.


%

\appendices
\section{Proof of the First Zonklar Equation}
Appendix one text goes here.

\section{}
Appendix two text goes here.

\section*{Acknowledgment}

The authors would like to thank...

\ifCLASSOPTIONcaptionsoff
  \newpage
\fi



\bibliographystyle{IEEEtran}
\bibliography{bibtex/bib/IEEE.bib}
\end{document}